\documentclass[%
 reprint,
 amsmath,amssymb,
 aps,
 prx,
 longbibliography,
]{revtex4-1}

\usepackage{graphicx}
\usepackage{dcolumn}
\usepackage{bm}
\usepackage{bbm}

\usepackage{algorithm, algorithmic}
\usepackage{setspace}
\usepackage{subcaption}

\begin{document}

\preprint{APS/123-QED}

\title{Computational landscape of user behavior on social media}

\author{David Darmon}
\email{ddarmon@monmouth.edu}
\affiliation{Department of Mathematics, Monmouth University, West Long Branch, New Jersey 07764, USA}
\author{William Rand}
\affiliation{Department of Business Management, North Carolina State University, Raleigh, North Carolina 27695, USA}
\author{Michelle Girvan}
\affiliation{Department of Physics, University of Maryland, College Park, Maryland 20742, USA}
\affiliation{Santa Fe Institute, 1399 Hyde Park Road, Santa Fe, NM 87501, USA}
\affiliation{London Mathematical Laboratory, 8 Margravine Gardens, W6 8RH London, UK}

\date{\today}

\begin{abstract}
With the increasing abundance of `digital footprints' left by human interactions in online environments, \emph{e.g.}, social media and app use, the ability to model complex human behavior has become increasingly possible. Many approaches have been proposed, however, most previous model frameworks are fairly restrictive. We introduce a new social modeling approach that enables the creation of models directly from data with minimal \emph{a priori} restrictions on the model class. In particular, we infer the minimally complex, maximally predictive representation of an individual's behavior when viewed in isolation and as driven by a social input. We then apply this framework to a heterogeneous catalog of human behavior collected from fifteen thousand users on the microblogging platform Twitter. The models allow us to describe how a user processes their past behavior and their social inputs. Despite the diversity of observed user behavior, most models inferred fall into a small subclass of all possible finite-state processes. Thus, our work demonstrates that user behavior, while quite complex, belies simple underlying computational structures.
\end{abstract}

\pacs{Valid PACS appear here}
\maketitle


\section{Introduction}


The current decade has been marked by an increasing availability of high-resolution, heterogeneous data sets capturing human behavior in both real-world and digital environments~\cite{eubank2004structural,compton2014geotagging,kramer2014experimental,toole2015coupling}. This has made possible, for the first time, large scale investigations into human behavior across diverse groups of individuals. Of such phenomena, human communication patterns are one of the most well-studied. Such studies have included written correspondences~\cite{oliveira2005human}, email correspondences~\cite{malmgren2008poissonian,malmgren2009characterizing}, and call/SMS records~\cite{jiang2013calling,wu2010evidence}. The characteristics of these behavioral patterns include heavy tails, seasonality, and burstiness. This is certainly still an active field of research, and many authors have called into question whether the observed patterns are truly universal characteristics of human behavior or epiphenomena of the methods used in data collection and analysis~\cite{goh2008burstiness,kivela2014estimating,ross2015understanding}.






The standard model for human communication patterns treats the observed behavior as a realization from some sort of point process. Typically, the point process is taken to be a renewal process, where the observed behavior is completely specified by a distribution over the times between activity. To account for the complex properties of human behavior enumerated above, the interevent distribution is specified to have a heavy right tail, which naturally gives rise to burstiness. The authors of~\cite{malmgren2008poissonian,malmgren2009characterizing} develop a refinement of this model which incorporates seasonality by allowing an individual to pass between passive and active states, where the behavior within the active state is governed by a Poisson process. Further refinements of this model allow the activity during the active periods to follow non-Poissonian dynamics~\cite{ross2015understanding}.



We undertake an analysis of human communication that does not \emph{a priori} assume a renewal or renewal-like model of the observed behavior. Motivated by the field of computational mechanics~\cite{shalizi2001computational}, we define our models explicitly in terms of a \emph{predictive} representation of the observed behavior. Unlike renewal process models, we do not assume the behavior of individuals only depends on the time between actions. We seek to understand the behavior locally in time, where locality is defined around periods of activity. Moreover, we explicitly incorporate the interactive aspect of online social media services, something missing from much of the work on modeling human interevent distributions, with~\cite{raghavan2013modeling} as a notable exception.

In~\cite{johnson2012enumerating}, the authors set out to elucidate the structural properties of stochastic processes using tools from computational mechanics. To do so, they restricted their investigation to the subset of stochastic processes that are \emph{finitary}, that is, those stochastic processes that have a representation with a finite number of ``causal'' states~\cite{wiesner2008computation}, as defined in Section~\ref{sec:self-driven}. In this work, instead of elucidating all possible finitary models, we approach the problem from the opposite direction: we seek to trace the computational landscape of human behavior in digital environments by discovering the finitary models present in user behavior, and then investigate their computational structure.

\begin{figure}[!h]
	\centering
	\includegraphics[width=0.5\textwidth]{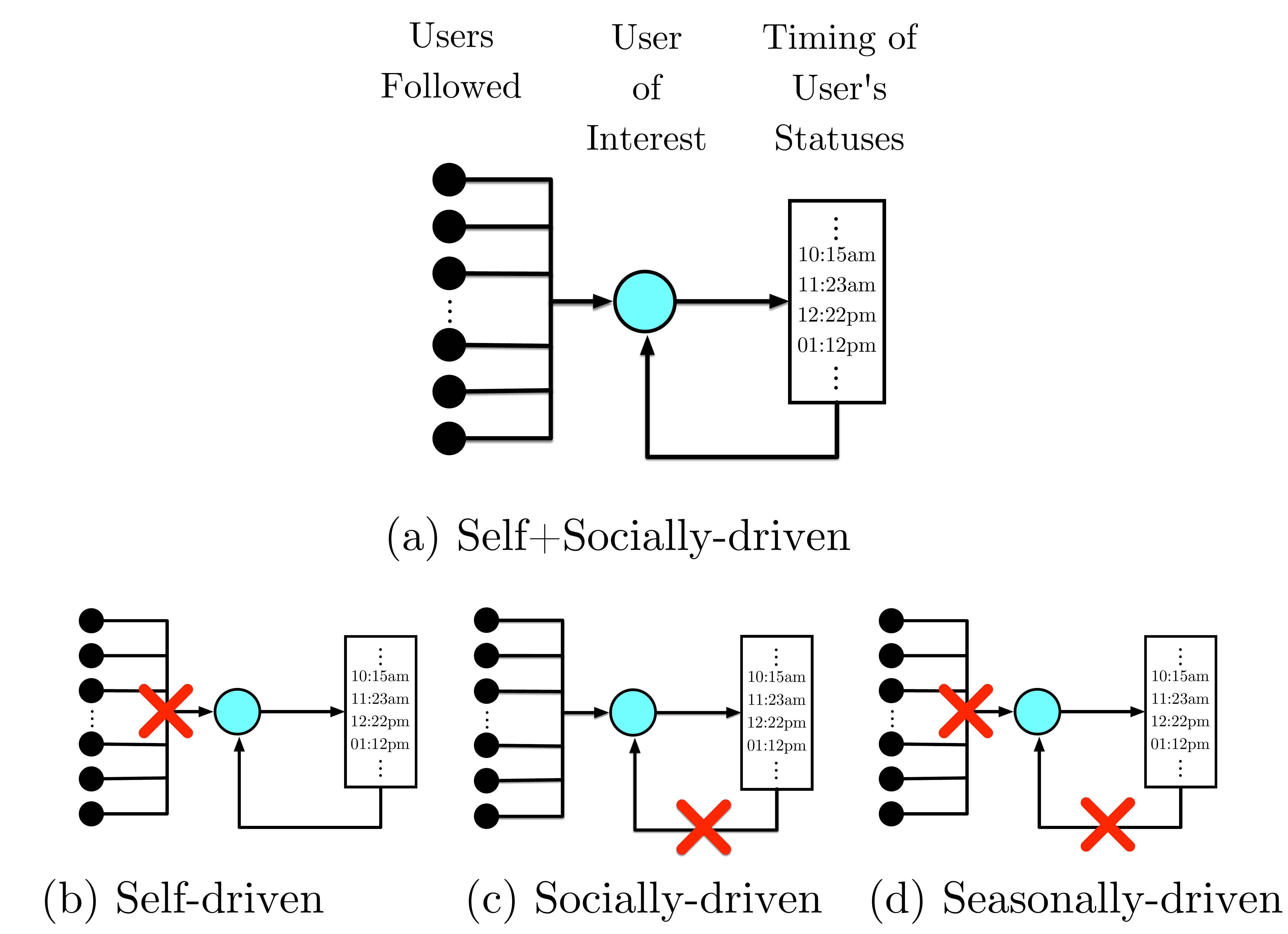}
	\caption{A schematic representation of the classes of models that we consider in this paper. (a) The most general case, where the user's observed behavior is influenced by their social inputs and their own past behavior. (b) The self-driven case, where the user's behavior only depends on their past behavior. (c) The socially-driven case, where the user's behavior only depends on their social inputs. (d) The seasonally-driven case, where the user's behavior can largely be attributed to time.}
	\label{Fig-schematic-models}
\end{figure}

We consider four models for user behavior on social media. Figure~\ref{Fig-schematic-models} provides a schematic representation of these models. The most general model (a) assumes that a user's future behavior is influenced by both their past behavior and the past behavior of their social network, which we call the self+socially-driven model. Models (b) and (c) are two restrictions of this model, the former where we assume that user's future behavior is only influenced by their past behavior, and the latter where we assume that the user's future behavior is only influenced by the past behavior of their social network. Finally, model (d) corresponds to the case where the user's behavior is entirely explained by time-of-day and day-of-week (\emph{i.e.} seasonality).

In the rest of the paper, we proceed as follows. In Section~\ref{sec:methods} we motivate and develop the four models just presented, and propose methods for inferring them. In Section~\ref{sec:descriptive-performance} we explore the descriptive performance of these models on a real world data set derived from 15K users on the microblogging platform Twitter. In Section~\ref{sec:causal-architectures}, we investigate the structure of the models present amongst the users in our data set, and discuss the implications of these models. Finally, we conclude with the implications of our present work for the study of human communication patterns.

\section{Methodology}

\label{sec:methods}


\subsection{User Behavior as a Discrete-Time Point Process}


Consider the behavior of a user on a social media service. At any given time instant, a user either posts to the social media service or not. Thus, the user's behavior may be modeled as a point process, where events correspond to posts. A naive model of the user's behavior might assume that they are equally likely to use the service during any time instant. Under this model, the time between uses is exponentially distributed, and their activity pattern would correspond to a realization from a Poisson process. However, human communication patterns are known to be exhibit non-trivial complexities not accounted for by this model~\cite{goh2008burstiness,rybski2012communication}, and thus more flexible models are required. In the following sections, we present three models that capture the observed complexity of human behavior in very different ways: a seasonally-driven model where the user's behavior is accounted for by time-of-day; a self-driven model where a user's behavior results from self-feedback; and a socially-driven model where a user's behavior results from both social- and self-feedback.

In practice, the information about human behavior on digital services is reported in seconds. Because we are interested in human-scale interactions between a user, their inputs, and the social media service, this time resolution is too fine grained. We begin by discretizing time into intervals of length $\delta$. We then ask if, during an interval $[(t-1)\delta, t \delta)$, the user was active. We denote this value for a user $v$ by $X_{t}(v)$ and define
\begin{align}
	X_{t}(v) = \left\{\begin{array}{ll}
	1 &: \text{user $v$ active during $[(t-1)\delta, t \delta)$} \\
	0 &: \text{otherwise}
	\end{array}\right. .
\end{align}
The choice of $\delta$ specifies the time scale of interest. For example, if we take $\delta = 1 \text{ day}$, then the process $\{X_{t}(v)\}$ captures the weekly patterns of behavior that the user exhibits. If instead we take $\delta = 1 \text{ hour}$, then $\{X_{t}(v)\}$ captures the daily patterns of the user. In this paper, we will take $\delta = 10 \text{ minutes}$, because we are interested in the short-timescale behavior of user behavior and user-user interaction. However, it is important to note that there is no single `correct' resolution when considering the behavior of a point process, and a multi-timescale analysis may be appropriate~\cite{marzen2015time}. Moreover, different time resolutions may be more or less appropriate for different users. Figure~\ref{Fig-user_seasonalities}a demonstrates the activity patterns $\{X_{t}(v)\}$ of three users at the 10 minute resolution, represented as a rastergram. Each row of the rastergram corresponds to a single day of activity, and each column of the rastergram corresponds to a ten minute window within a single day. A point occurs in the rastergram when $X_{t}(v) = 1$ for that day and time.

In a social media setting, a user has access to information provided by other users on the service. For example, a user might passively examine the messages generated by other users they follow, observe a particular form of communication directed at them, or actively investigate a keyword or topic. Generically, we will denote the inputs to a user as $Y_{t}(v)$. We will assume that the inputs to the user can be mapped to a finite alphabet $\mathcal{Y}$. As an example, if we consider $Y_{t}(v)$ to correspond to whether or not the user $v$ receives a mention during the time interval $[(t-1)\delta, t \delta)$, then we take $\mathcal{Y} = \{0 ,1\}$, where $y = 0$ corresponds to no mention during that time interval, and $y = 1$ corresponds to one or more mentions.

Our goal in this paper is the develop several contrasting models of a user's observed behavior $\{X_{t}(v)\}$.
We take a predictive view of modeling, where we seek to infer the probability that the user engages with the social media service, given their past history of engagement and the past history of their inputs. Let $X_{-\infty}^{t-1}(v) = (\ldots, X_{t-2}(v), X_{t-1}(v))$ be the past behavior of user $v$, and let $X_{t}^{\infty}(v) = (X_{t}(v), X_{t+1}(v), \ldots)$ be the future behavior of the user. Similarly, let $Y_{-\infty}^{t-1}(v) = (\ldots, Y_{t-2}(v), Y_{t-1}(v))$ and $Y_{t}^{\infty}(v) = (Y_{t}(v), Y_{t+1}(v), \ldots)$ be the past and future values of the user's inputs, considering $t$ as the present. Then we are interested in determining
\begin{align}
	P(X_{t}^{\infty}(v) \mid X_{-\infty}^{t-1}(v) = x_{-\infty}^{t-1}, Y_{-\infty}^{t-1} = y_{-\infty}^{t-1}(v)),
\end{align}
the distribution over user $v$'s behavior starting from time $t$, given their own past behavior and the past behavior of their inputs. For ease of presentation, in the following sections, we drop the dependence on $v$ in the notation, but emphasize that for each user $v$, we assume a unique model for that user's behavior.

\subsection{Seasonally-Driven Model: Inhomogenous Bernoulli}

\label{sec:model-seasonal}

\begin{figure}[!h]
	\centering
	\includegraphics[width=0.5\textwidth]{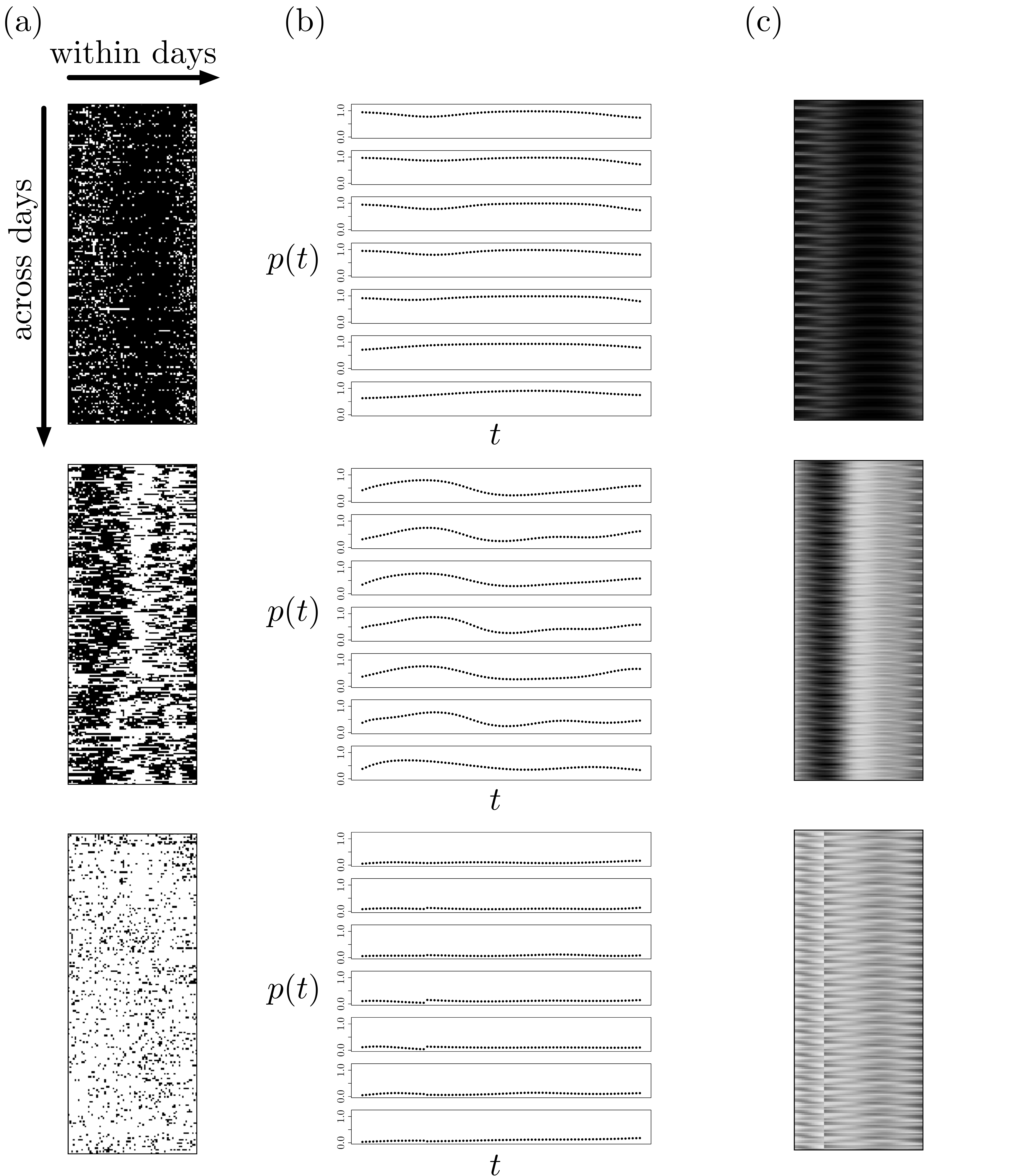}
	\caption{(a) Rastergram representation of the activity of three users on Twitter over a 44 week period. (b) The expected activity of the same three users. Each panel corresponds to the expected activity by day-of-week, from Monday to Sunday. (c) The expected activity from (b), laid out in the same format as the rastergram. Note that the color scale for each panel is taken from 0 to $\max\limits_{t} p_{v}(t)$ for each user $v$ to make the seasonality in the activity patterns more obvious.}
	\label{Fig-user_seasonalities}
\end{figure}

Thus far, we have specified our model of human behavior in terms of a discrete-time point process: the observed behavior of the user is either active or quiescent during any given interval of time. One of the simplest models that can capture some of the complexity of human behavior is a renewal process~\cite{esteban2012analysis,doerr2013lognormal}. From this perspective, the activity of the user is taken to occur at random times, with the time between occurrences governed by a distribution over the interarrival times. For example, if we take the interarrival distribution to be geometric with parameter $p$, then the renewal process is a Bernoulli process, the discrete-time analog of a Poisson process. Typically, the interarrival distribution is taken to have a long tail, to capture the fact that human behavior tends to be bursty, with long periods of quiescence punctuated by periods of high activity. See Figure~\ref{Fig-user_seasonalities} for examples of users who exhibit such behavior. Popular distributions for the interarrival times include log-normal, power law, and stretched exponential distributions~\cite{goh2008burstiness}.

Due to the inherent seasonality in human behavior, time-homogeneous renewal process-based models are almost certainly misspecified. For example, a typical user on Twitter will be more likely to be active during the daylight hours in their geographic area than during the nighttime hours. This fact may explain the long tails typically observed in studies of the activity patterns of humans~\cite{jo2012circadian}. Moreover, we see such daily and weekly seasonality patterns in the aggregate behavior of users on Twitter. Because of this, we consider a time-inhomogeneous point process model for a user's observed activity, where the probability a user is active during any time interval is independent of their previous activity and the activity of their inputs, and varies smoothly with time,
\begin{align}
	P(X_{t} = 1 \mid X_{-\infty}^{t-1} = x_{-\infty}^{t-1}, Y_{-\infty}^{t-1} = y_{-\infty}^{t-1}, ) = p(t) \label{eqn:instantiation-seasonally-driven}
\end{align}
Moreover, we assume that $p(t)$ is periodic, $p(t + \tau) = p(t)$, with $\tau$  chosen such that for a coarsening interval $\delta$, $\tau \delta = 1$ week. We take (\ref{eqn:instantiation-seasonally-driven}) as our instantiation of the seasonally-driven model from Figure~\ref{Fig-schematic-models}d.

We estimate the individual seasonality $p(t)$ for each user via a Generalized Additive Model (GAM)~\cite{hastie1990generalized}
\begin{align}
	\text{logit}(p(t)) = \beta_{0} + f(t) \label{eqn:GAM-model}
\end{align}
where
\begin{align}
	\text{logit}(x) = \log \left( \frac{x}{1-x}\right)
\end{align}
using the \texttt{mgcv} package in R. Figure~\ref{Fig-user_seasonalities} demonstrates the observed behavior of several users, along with their estimated activity probabilities $p(t)$.

\subsection{Self-driven Model: The $\epsilon$-machine}

\label{sec:self-driven}


The previous model assumes that the user's activity during any time interval is independent of their activity during other time intervals, and accounts for the seasonality and bursting observed in a user by allowing the probability of their activity to vary according to time-of-day and day-of-week. Alternatively, a user might exhibit burstiness due to self-excitation. As an example, the user might be isolated from the devices they use to interact with the social media service, which would lead to a period of quiescence. Then, upon regaining access to their devices, they might use the service, which could lead to a self-excitation to continue using the service.


This sort of behavior motivates an autoregressive model for the user's behavior, where their behavior in the future is determined by their past behavior. That is, the probability that they behave a certain way in the future starting at time $t$ is determined by how they behaved up until time $t$ and does not depend on their inputs,
\begin{align}
\begin{split}
P(X_{1}^{\infty} \mid X_{-\infty}^{0} = x_{-\infty}^{0}). \label{self-driven-prob}
\end{split}
\end{align}
This model assumes that the users behavior is conditionally stationary~\cite{caires2005non}. For human behavior, this assumption may not hold in general, and thus care must be taken in applying this model with actual data. We address this in Section~\ref{data-preprocessing}, where we specify our procedure for day-casting user behavior. Previous work has found this model to perform well with many users on Twitter~\cite{ver2012information,raghavan2013modeling,darmon2013predictability}.

\begin{figure}
        \centering
        \begin{subfigure}[b]{0.225\textwidth}
                \includegraphics[width=\textwidth]{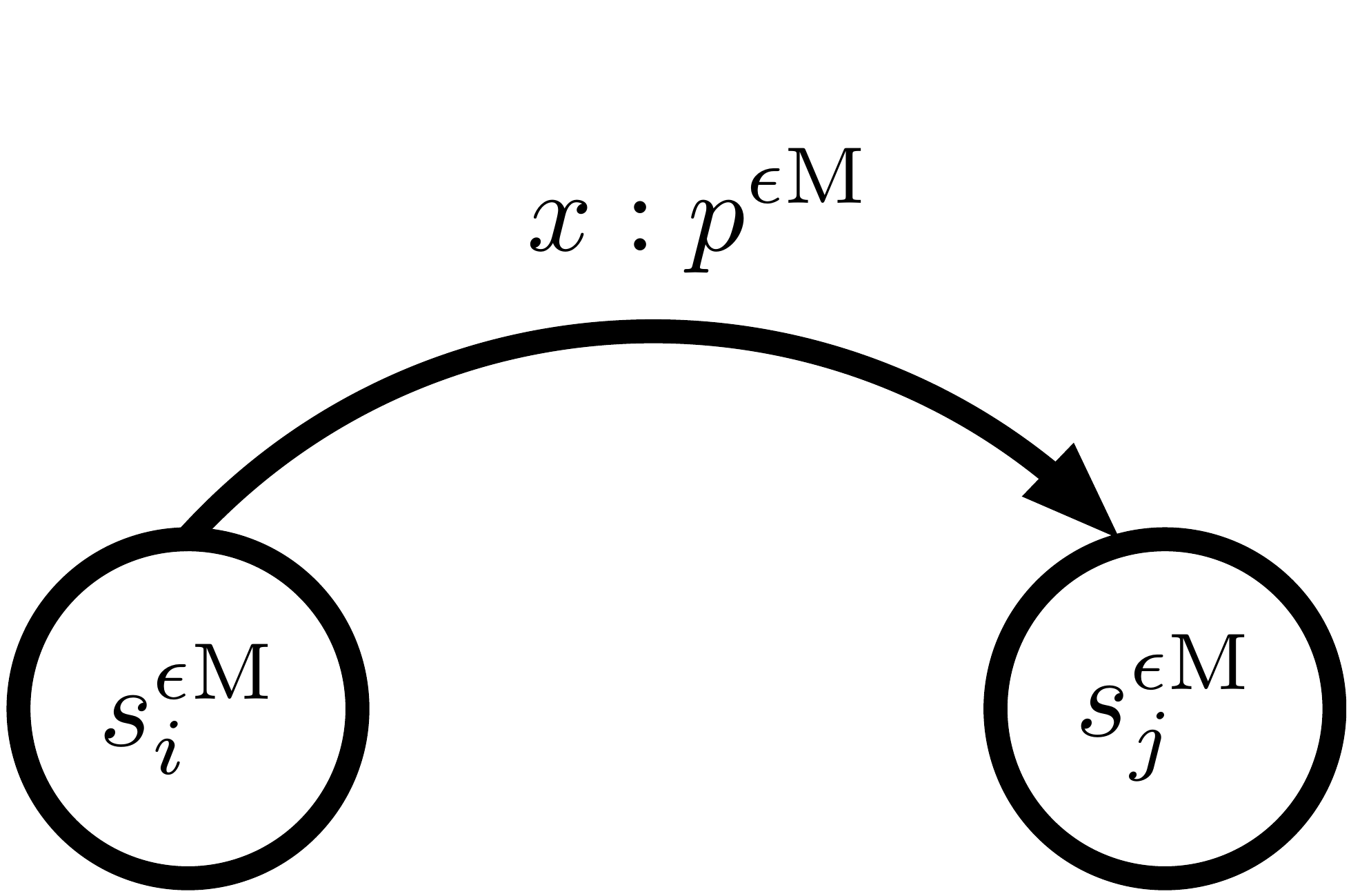}
                \caption{$\epsilon$-machine}
                \label{Fig:eM-demo}
        \end{subfigure}
        \begin{subfigure}[b]{0.225\textwidth}
                \includegraphics[width=\textwidth]{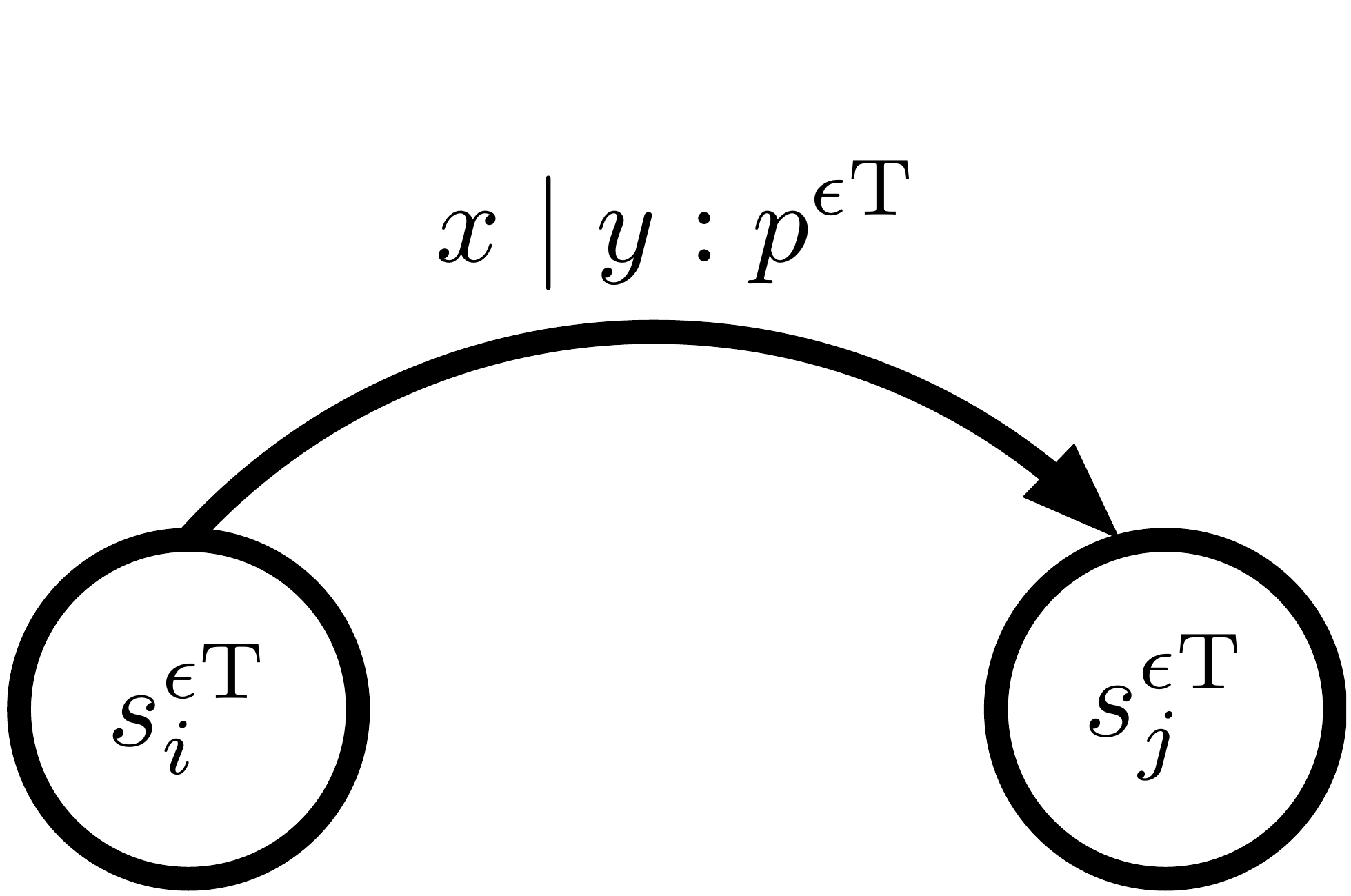}
                \caption{$\epsilon$-transducer}
                \label{Fig:eT-demo}
        \end{subfigure}
        \label{Fig:eM+eT-demo}
        \caption{Transitions between (a) $\epsilon$-machine and (b) $\epsilon$-transducer causal states. Each transition is labeled by the (a) marginal and (b) joint emission symbol, as well as the transition probability $p^{\epsilon\text{M/T}}$.}
\end{figure}

We take the self-driven model from Figure~\ref{Fig-schematic-models}b to correspond to a stochastic process governed by~(\ref{self-driven-prob}), which we develop using computational mechanics~\cite{shalizi2001computational}. Computational mechanics provides the unique, minimally complex, maximally predictive representation of a discrete state, discrete time stochastic process $\{ X_{t}\}_{t \in \mathbb{Z}}$ over the alphabet $\mathcal{X}$. The insight of computational mechanics is that when considering the predictive distribution (\ref{self-driven-prob}), it is typically more useful to consider a \emph{statistic} of the past $x_{-\infty}^{t-1}$ rather than the entire past itself. It can be shown that the unique minimal sufficient predictive statistic of the past $X_{-\infty}^{t-1}$ for the future $X_{t}^{\infty}$ of a conditionally stationary stochastic process is the equivalence class over predictive distributions. That is, two pasts $\hat{x}_{-\infty}^{t-1}$ and $\tilde{x}_{-\infty}^{t-1}$ are considered equivalent $\hat{x}_{-\infty}^{t-1} \sim_{\epsilon} \tilde{x}_{-\infty}^{t-1}$ if and only if $P(X_{t}^{\infty} \mid X_{-\infty}^{t-1} = \hat{x}_{-\infty}^{t-1}) = P(X_{t}^{\infty} \mid X_{-\infty}^{t-1} = \tilde{x}_{-\infty}^{t-1})$. The equivalence relation induces a set of equivalence classes $\mathcal{S}$ called the causal states of the process. The causal states, the allowed transitions between them, and the probability associated with the transitions is called the $\epsilon$-machine for the process $\{X_{t}\}_{t \in \mathbb{Z}}$. For a stochastic process with a finite number of predictive equivalence classes, the $\epsilon$-machine may be represented as deterministic finite automata, where the states of the automata correspond to the causal states, and the transitions between states are determined by the outputs $x \in \mathcal{X}$. A demonstration of a portion of such a representation is given in Figure~\ref{Fig:eM-demo}.

\subsection{Socially-driven Models: The $\epsilon$-transducer}

\label{sec-model-social}

The previous two models assume that either the user is driven seasonally by time-of-day type influences, or that the user is self-driven. On a social media service, we expect a user to interact with other users, and therefore we would expect the user's behavior to be associated with the behavior of those users. For example, a user might become more likely to tweet if they have recently been mentioned by another user. Such social associations are captured by the social inputs $\{Y_{t}(v)\}_{t \in \mathbb{Z}}$ of a user $v$. In particular, we will focus on the mention history of the user,
\begin{align}
	Y_{t}(v) = \left\{\begin{array}{ll}
	1 & : \text{mention of $v$ in $[(t-1)\delta, t \delta)$} \\
	0 & : \text{otherwise}
	\end{array}\right. .
\end{align}
That is, $Y_{t}(v)$ corresponds to whether or not the user $v$ received any mentions during the time window of length $\delta$ indexed by $t$.

We take the modeling perspective where the user acts as a \emph{transducer}, mapping their own past behavior and the past behavior of their social inputs into their future behavior. More explicitly, as with the self-driven example, we seek the minimally complex, maximally predictive model for the user's behavior. Again, computational mechanics provides such a model via the $\epsilon$-transducer~\cite{crutchfield1994optimal,shalizi2001causal}. The theory for the $\epsilon$-transducer has been recently developed in~\cite{barnett2014computational}. The main insight is the same as for the $\epsilon$-machine: we define an equivalence relation over joint input-output pasts such that two pasts are equivalent if they induce the same predictive distribution over the future output. As with the $\epsilon$-machine, this equivalence relation induces a partition of the joint input-output past into channel causal states. Transitions between channel causal states occur on joint input-output pairs, and thus an $\epsilon$-transducer with finitely many channel causal states, like an $\epsilon$-machine, can be represented as a deterministic automata. The representation is given in Figure~\ref{Fig:eT-demo}.

For the socially-driven model, we distinguish between the self-memoryful $\epsilon$-transducer and self-memoryless cases, corresponding to the models from Figure~\ref{Fig-schematic-models}a and Figure~\ref{Fig-schematic-models}b, respectively. For the self-memoryful $\epsilon$-transducer, we construct the equivalence classes using the joint input-output pasts, while for the self-memoryless $\epsilon$-transducer, we construct the equivalence classes using only the output pasts. Thus, the self-memoryless $\epsilon$-transducer assumes that a user's behavior is purely driven by their social input. We consider both cases under the heading of socially-driven models.

\subsection{Data Collection and Pre-processing}

\label{data-preprocessing}

The activity of the 15K users was collected over a 49 week period, from 6 June 2014 to 15 May 2015. After data cleaning to account for outages in the data collection, 44 weeks of data were generated. 
We did not include the quiescent users in our analysis. As described above, the self-driven and socially-driven models assume that a user's behavior can be modeled as a conditionally stationary stochastic process, where the distribution over futures is independent of the time index conditional on the observed past of the user or the observed joint input / output past, respectively. In order to make this assumption approximately true, we `daycast' the time series associated with each user as follows. For a user, we determine their time zone as recorded by Twitter, and window their activity to be between 9 AM and 10 PM during their local time. We take this time window to capture the waking hours of a typical individual.

For this study, we split the 44 weeks of data into 28 weeks of training data and 16 weeks of testing data. The training data is used to select and infer the models, as we describe in the next section. The testing data is used for the comparison of these models in terms of their predictive and descriptive performance. This train/test split is performed to ensure that we obtain unbiased estimates of how the models perform for each user.


\subsection{Model Inference and Selection}

\label{sec:mod-inf}


For the seasonally-driven model, the only model parameter associated with each user is the smoothing parameter for the splines used to estimate the non-parametric term $f(t)$ in (\ref{eqn:GAM-model}). This parameter is chosen using generalized cross validation~\cite{hastie2009elements} on the 28 weeks of training data.



For $\epsilon$-machine reconstruction, we use the Causal State Splitting Reconstruction (\texttt{CSSR}) algorithm~\cite{CSSR-UAI-2004} to infer the models from data. For $\epsilon$-transducer reconstruction, we use the Transducer Causal State Splitting Reconstruction (\texttt{transCSSR}) algorithm, described in Appendix~\ref{transCSSR-algorithm}. Both \texttt{CSSR} and \texttt{transCSSR} require the specification of a tuning parameter $\alpha$ that controls the probability of splitting histories from a state when no such split should occur, and $L_{\max}$, the maximum history length used in determining the candidate causal / transducer states. We fix $\alpha$ at 0.001. The maximum history length $L_{\max}$ directly balances between the flexibility of the model and the precision with which the probabilities may be estimated. As an example, suppose a maximum history length $L_{\max}$ is sufficient to resolve the causal states. In the extreme case that each history of length $L_{\max}$ specifies a unique predictive distribution (an order $L_{\max}$ Markov model), then the model would result in $|\mathcal{X}|^{L_{\max}}$ causal states. However, as we increase $L_{\max}$, we also necessarily decrease the number of examples of each history used to estimate the predictive distribution. This can result in spurious splitting of histories.


We use $K$-fold cross-validation~\cite{hastie2009elements} to choose the appropriate $L_{\max}$ for each user. In particular, for each user, we randomly partition the 196 days in the training set into $K = 5$ folds. For a held out fold $k$ whose time indices are $\mathcal{T}_{k}$, define the empirical total variation (ETV) distance between their observed behavior during that fold and the model inferred using the $K - 1$ remaining folds as
\begin{align}
	\begin{split}
	&\text{ETV}(L_{\max}, k) \\&= \frac{1}{|\mathcal{T}_{k}|} \sum_{t \in \mathcal{T}_{k}} \left(\frac{1}{2} \sum_{x \in \mathcal{X}} \big|\delta_{X_{v}(t),x} - p_{v}^{(L_{\max}, -k)}(x, t)\big|\right) \label{eqn-ETVD}
	\end{split}
\end{align}
where $\delta_{x', x}$ is the Kronecker delta and $p_{v}^{(L_{\max}, -k)}(x, t)$ is the probability of observing outcome $x$ at time $t$ using the model inferred with all of the data except that from $k$th fold. In the binary case, (\ref{eqn-ETVD}) reduces to
\begin{align}
	\begin{split}
	&\text{ETV}(L_{\max}, k) \\&= \frac{1}{|\mathcal{T}_{k}|} \sum_{t \in \mathcal{T}_{k}}\big|\delta_{X_{v}(t), 1} - p_{v}^{(L_{\max}, -k)}(1, t)\big| \label{eqn-ETVD-simple}
	\end{split}
\end{align}
Thus, we see that (\ref{eqn-ETVD}) quantifies the model performance by comparing the actual outcome for the user to the estimated probability of that outcome using the model. We can then compute the average of the empirical total variation over the held out sets,
\begin{align}
	\text{ETV}(L_{\max}) = \frac{1}{K} \sum_{k = 1}^{K} \text{ETV}(L_{\max}, k),
\end{align}
and choose $L_{\max}$ to minimize this value. We perform this optimization using $L_{\max}$ from 1 to 6, which for $\delta = 10 \text{ minutes}$ corresponds to a time span between ten minutes and an hour.




%






\section{Results}

\subsection{Descriptive Performance Across the Model Classes}

\label{sec:descriptive-performance}

We begin by examining the ability of the four models developed in Sections~\ref{sec:model-seasonal}--\ref{sec-model-social} to describe a given user's behavior. To do so, we compute the ETV, as defined by~(\ref{eqn-ETVD}), between the held out test data and the cross-validated models of each type. This provides us with a measure of how the models generalize to unseen behavior, and thus an indication of how well the models describe a user's behavior. Because the ETV for a given user depends on their overall activity level, we standardize the ETV for a model  $\mathcal{M}$ by the ETV for the seasonality model, giving us a score function
\begin{align}
	\begin{split}
	\text{Score}(\mathcal{M}; S) = \frac{\text{ETV}(S)}{\text{ETV}(\mathcal{M})}. \\
	\end{split}
	\label{eqn-score}
\end{align}
Recalling that a smaller ETV value indicates a smaller distance between the observed behavior and the model predictions, we see that $\text{Score}(\mathcal{M}; S)$ will be greater than 1 when model $\mathcal{M}$ outperforms the seasonal model, and smaller than 1 otherwise.

\begin{figure}[!h]
	\centering
	\includegraphics[width=0.50\textwidth]{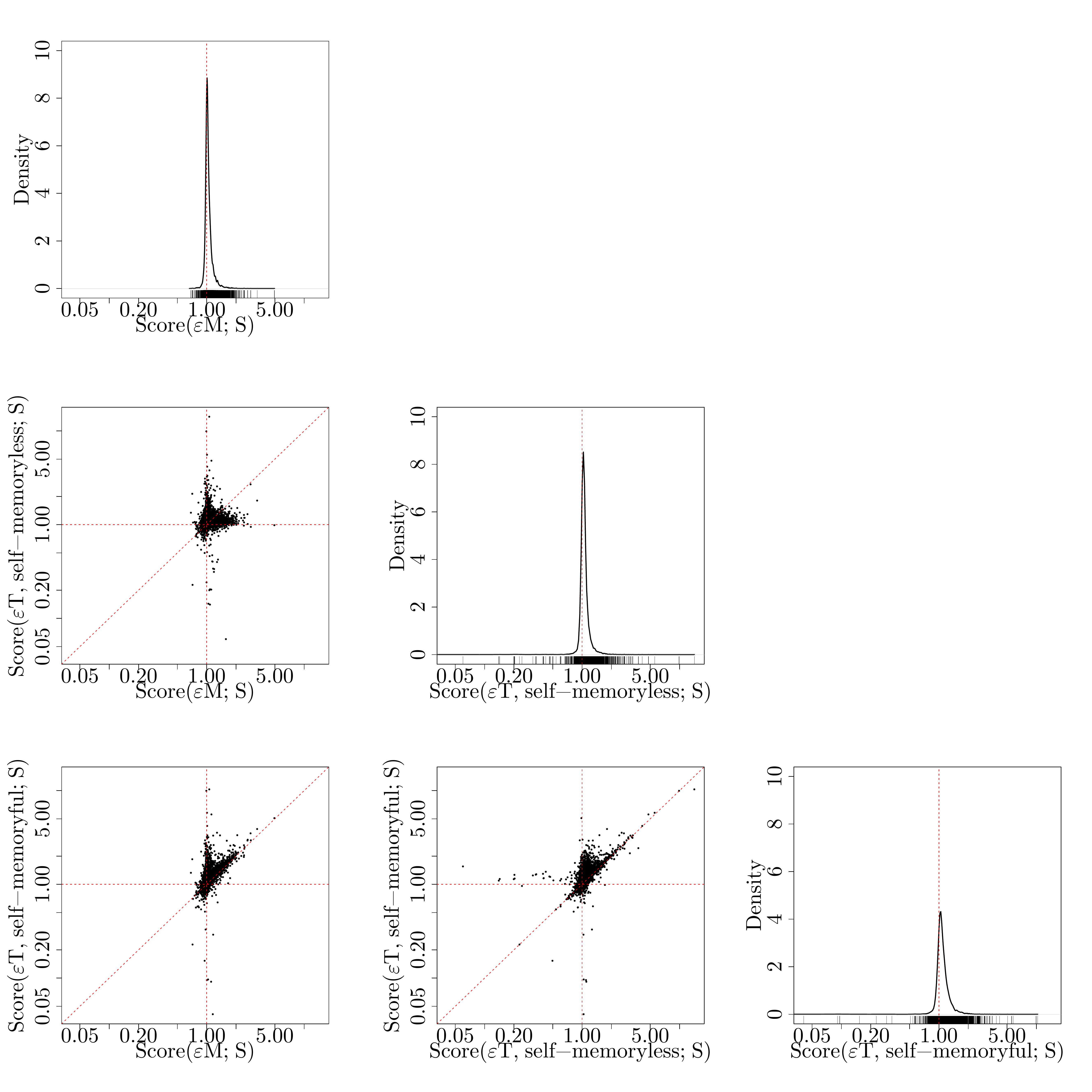}
	\caption{The descriptive performance of the self-driven and socially-driven models for each user on the test set data relative to the seasonally-driven model using the ETV-based score defined by (\ref{eqn-score}). The diagonal entries show the density of scores for the $\epsilon$-machine, self-memoryless $\epsilon$-transducer, and self-memoryful $\epsilon$-transducer across the users. The off-diagonal entries compare the scores between the non-seasonal models. A score greater than 1 indicates that the referenced model outperformed the seasonally-driven model.}
	\label{Fig-score-score-all}
\end{figure}

The scores across all users for all models are shown in Figure~\ref{Fig-score-score-all}. The diagonal shows the densities of scores across the users for each model type. The self- and socially-driven models generally perform better than the seasonal model, with all of the score densities having a heavy tail right tail. We see that the self-memoryful $\epsilon$-transducer performs best, with a score greater than 1 for 82.1\% of the users. The self-memoryless $\epsilon$-transducer is next best, with a score greater than 1 for 79.4\% of the users. The $\epsilon$-machine has a score greater than 1 for 72\% of the users.

We further summarize the pairwise comparisons between the non-seasonal models in Table~\ref{tab:model-performance}. As expected, the self-memoryful $\epsilon$-transducer outperforms both the $\epsilon$-machine and the self-memoryless $\epsilon$-transducer on most users, as indicated by the mass of users above the identity line in the first quadrant in the bottom row of Figure~\ref{Fig-score-score-all}. However, the users are much more equally split between those where the $\epsilon$-machine outperforms the self-memoryless $\epsilon$-transducer (45.2\%) and \emph{vice versa} (54.8\%).


\begin{table}
	\centering
	\caption{Pairwise comparison between the $\epsilon$-machine ($\epsilon$M), self-memoryless $\epsilon$-transducer ($\epsilon$T-ML), and self-memoryful $\epsilon$-transducer ($\epsilon$T-MF) across the users. Each entry in the table gives the proportion of users with $\text{Score}(\mathcal{M}_{1}; S) > \text{Score}(\mathcal{M}_{2}; S)$.}
\begin{tabular}{l | c c c}
   $\mathcal{M}_{1}$ \textbackslash \, $\mathcal{M}_{2}$ & $\epsilon$M & $\epsilon$T-ML & $\epsilon$T-MF\\
  \hline
  $\epsilon$M & --- & 0.452 & 0.244 \\
  $\epsilon$T-ML & 0.548 & --- & 0.336 \\
  $\epsilon$T-MF & 0.756 & 0.618 & --- \\
\end{tabular}
\label{tab:model-performance}
\end{table}

\subsection{$\epsilon$-machine Causal Architectures}

\label{sec:causal-architectures}


We next explore the typical $\epsilon$-machine architectures across the users. The number of causal states for an $\epsilon$-machine gives a rough indication of the complexity of the user's behavior since each causal state indicates a further refinement of the past necessary for predictive sufficiency. In fact, the logarithm of the number of states is called the \emph{topological complexity} of the $\epsilon$-machine~\cite{crutchfield1994calculi}. We find that most users are best described by models with a small number of states, with 95\% of users having 13 or fewer causal states, and the largest $\epsilon$-machine having 58 states. Note that the maximum number of possible causal states when $L_{\max} = 6$ is $2^{6} = 64$.

\begin{figure}[!h]
	\centering
	\includegraphics[width=0.2\textwidth]{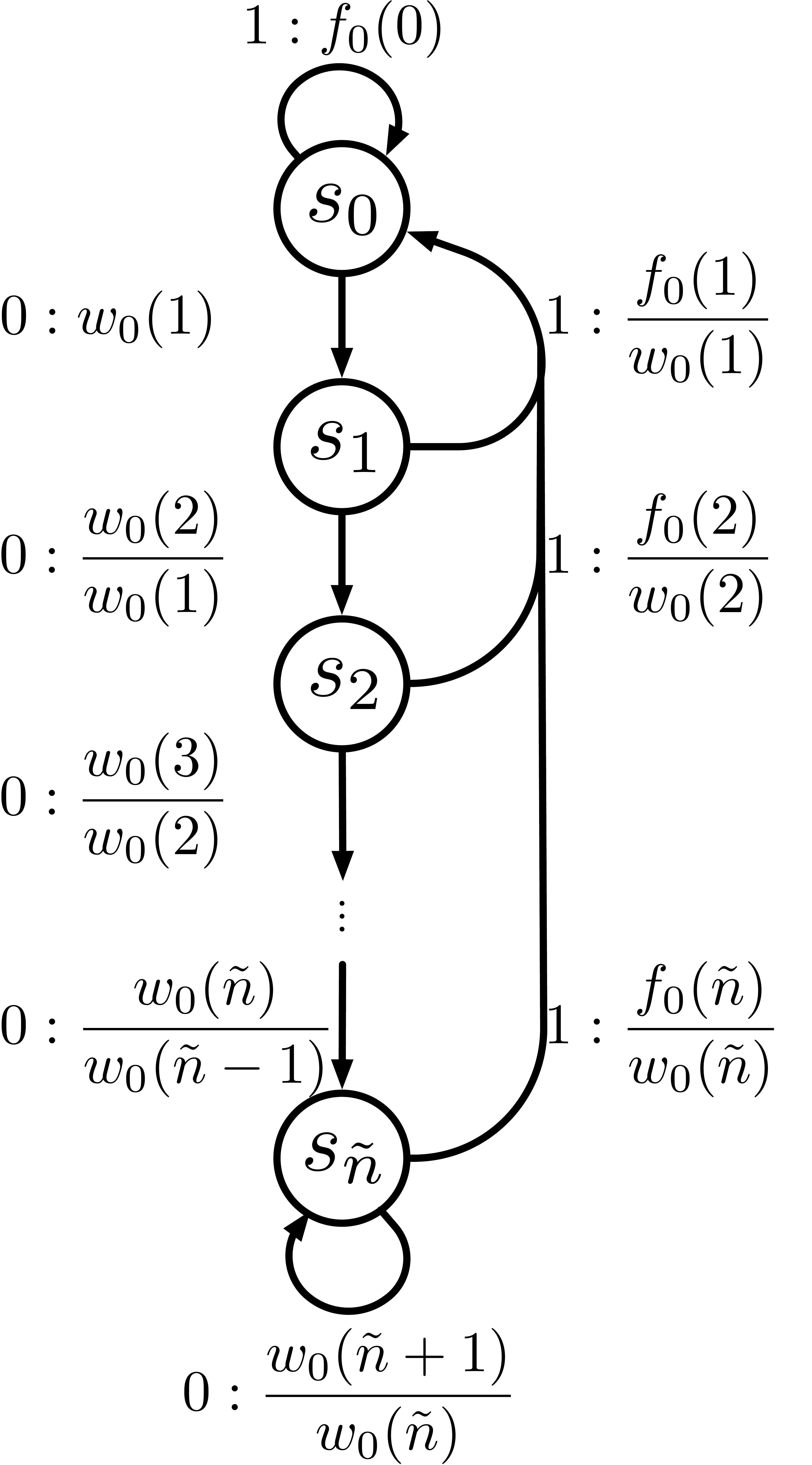}
	\includegraphics[width=0.2\textwidth]{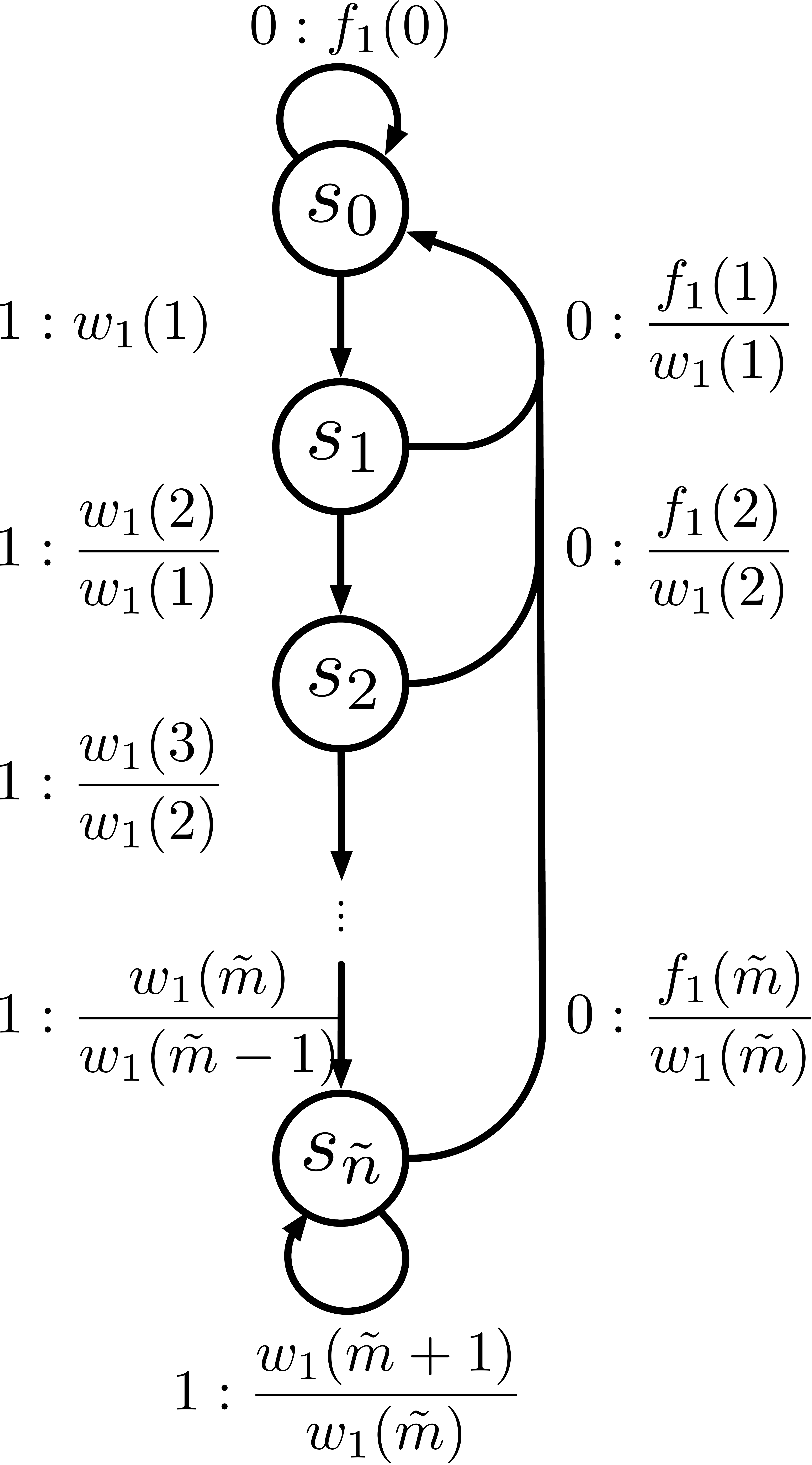}
	\caption{The $\epsilon$-machine representations of a renewal process of the eventually $\Delta_{0}$-Poisson type with characteristic $(\tilde{n}, \Delta_{0} = 1)$ (left) and of the reverse eventually $\Delta_{1}$-Poisson type with characteristic $(\tilde{m}, \Delta_{1} = 1)$ (right). 67.9\% of users have $\epsilon$-machines of the renewal-type, while only 0.59\% have $\epsilon$-machines of the reverse renewal-type.}
	\label{Fig-eM-renewal-architecture}
\end{figure}

\textbf{Renewal Process.} We next consider the general types of stochastic processes captured by many of the $\epsilon$-machines. We find that a large proportion of the users have $\epsilon$-machines which correspond to a generalization of a discrete-time renewal process. Recall that a discrete-time renewal process is a point process such that the lengths $\{N_{i}\}$ of periods of quiescence (runs of 0s between successive 1s) are independent and distributed according to an inter-arrival distribution $f_{0}(n) = P(N = n)$~\cite{marzen2014informational}. Equivalently, discrete-time renewal processes can be defined in terms of the survival function $w_{0}(n) = P(N \geq n)$. Because discrete-time renewal processes are a special case of the more general processes described by~(\ref{self-driven-prob}), their $\epsilon$-machine architecture takes on a very particular form~\cite{marzen2014informational}. The $\epsilon$-machine for a discrete-time renewal process has a unique start state transitioned to after a period of activity, and transitions after a period of quiescence  traverse a chain of states that counts the number of time points since an activity period occurred. We reproduce the generic architecture found amongst the renewal process $\epsilon$-machines in Figure~\ref{Fig-eM-renewal-architecture} (left). This is a special finite state case of the more general architecture for a discrete-time renewal process. In the nomenclature introduced in~\cite{marzen2014informational}, this is an $\tilde{n}$ eventually $\Delta_{0}$-Poisson process with characteristic parameters $(\tilde{n}, \Delta_{0} = 1)$, where $\tilde{n}$ refers to the number of quiescent time steps necessary for the $\epsilon$-machine to behave as a Poisson (Bernoulli) process, and the $\Delta_{0}$ refers to the smallest resolution at which the inter-event times may be coarse-grained and remain geometrically distributed. Such a process has an inter-event distribution
\begin{align}
f_{0}(n) = \left\{ \begin{array}{ll}
p_{0}(n) &: n = 0, \ldots, \tilde{n} \\
f_{0}(\tilde{n}) \lambda_{0}^{n - \tilde{n}} &: n > \tilde{n} \end{array}\right..
\end{align}
where $\{ p_{0}(n)\}_{n = 0}^{\tilde{n}}$ specify the initial $\tilde{n} + 1$ values of the inter-event distribution and $\lambda_{0} = \frac{1 - \sum_{n = 0}^{\tilde{n}}p_{0}(n)}{1 - \sum_{n = 0}^{\tilde{n-1}}p_{0}(n)}$. We note that using \texttt{CSSR} with finite $L_{\max}$ necessarily results in the reconstruction of finite state $\epsilon$-machines, and thus for  an $\tilde{n}$ eventually $\Delta_{0}$-Poisson processes with $\tilde{n} > L_{\max}$, the inferred $\epsilon$-machine will be an approximation to the longer memory process. In fact, this motivates a particular family of parametric models with parameters $\tilde{n}$ and $\{ p_{0}(n)\}_{n = 0}^{\tilde{n}}$ which specify the initial inter-event behavior. We emphasize that this particular family of parametric models was not assumed, but rather discovered via the use of \texttt{CSSR}.

\textbf{Reverse Renewal Process.} A renewal process is specified by a distribution $f_{0}(n)$ over run lengths $\{N_{i}\}$ of quiescence. For such a process, the distribution $f_{1}(m)$ over run lengths $\{M_{i}\}$ of activity follows a geometric distribution. One could also define a process where these roles are reversed: the  distribution $f_{1}(m)$ over run lengths of activity takes an arbitrary form, and the distribution $f_{0}(n)$ over run lengths of quiescence follows a geometric distribution. We call such a process a $\emph{reverse renewal process}$, since the roles of quiescence and activity are reversed. The $\epsilon$-machine for a reverse renewal process is given in Figure~\ref{Fig-eM-renewal-architecture} (right). In analogy to the $\tilde{n}$ eventually $\Delta_{0}$-Poisson process, we call this process a reverse $\tilde{m}$ eventually $\Delta_{1}$-Poisson process, which has the inter-quiescence distribution given by
\begin{align}
f_{1}(m) = \left\{ \begin{array}{ll}
p_{1}(m) &: m = 0, \ldots, \tilde{m} \\
f_{1}(\tilde{m}) \lambda_{1}^{m - \tilde{m}} &: m > \tilde{m} \end{array}\right.
\end{align}
where $\{ p_{1}(m)\}_{m = 0}^{\tilde{m}}$ specify the initial $\tilde{m} + 1$ values of the inter-quiescence distribution and $\lambda_{1} = \frac{1 - \sum_{m = 0}^{\tilde{m}}p_{1}(m)}{1 - \sum_{m = 0}^{\tilde{m-1}}p_{1}(m)}$.

\begin{figure}[!h]
	\centering
	\includegraphics[width=0.5\textwidth]{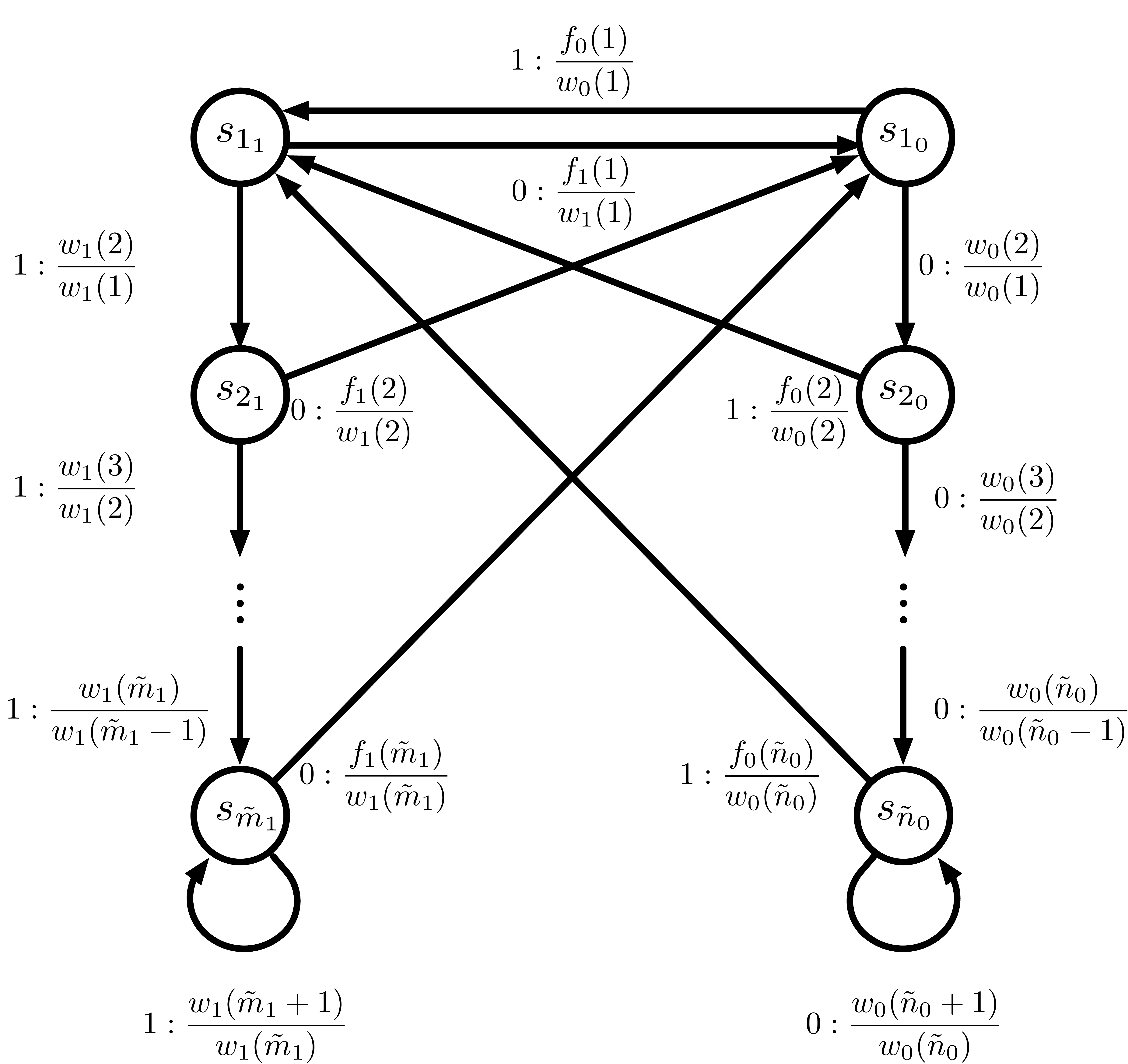}
	\caption{The $\epsilon$-machine representation of an alternating renewal process of the eventually $(\Delta_{1}, \Delta_{0})$-Poisson type with characteristic $(\tilde{m}_{1}, \tilde{n}_{0}, \Delta_{1} = 1, \Delta_{0} = 1)$. 8.7\% of users have $\epsilon$-machines of the alternating renewal type.}
	\label{Fig-eM-mixed-renewal-architecture}
\end{figure}

\textbf{Alternating Renewal Process.} More generally, we can define a class of processes such that the distributions over run lengths of activity \emph{and} quiescence are allowed to deviate from the geometric distribution. Such processes are known as \text{alternating renewal processes}. An alternating renewal process switches between periods of quiescence and activity with probabilities governed by the quiescence length $f_{0}(n)$ and activity length $f_{1}(m)$ distributions. The $\epsilon$-machine for an alternating renewal process with eventually geometric distributions for both inter-arrival and inter-quiescence is given in Figure~\ref{Fig-eM-mixed-renewal-architecture}. We call such a process an alternating $(\tilde{m}_{1}, \tilde{n}_{0})$ eventually $(\Delta_{1}, \Delta_{0})$-Poisson process. Again, this class of processes offers another parametric model for user behavior, with parameters $\tilde{n}_{0}, \tilde{m}_{1}, \{ p_{0}(n)\}_{n = 0}^{\tilde{n}},$ and $\{ p_{1}(m)\}_{m = 0}^{\tilde{m}}$.

Because of the stereotyped architecture of the $\epsilon$-machines for renewal, reverse renewal, and alternating renewal processes, we can easily identify those users whose $\epsilon$-machines have these architectures. An $\epsilon$-machine represents an alternating renewal process if and only if there is precisely one state transitioned to on an $x$ from a state transitioned to on an $x' \neq x$, $x \in \{0, 1\}$. For example, the state transitioned to on a 1 from states transitioned to on a 0 represents the start of a run of 0s. The $\epsilon$-machines for renewal / reverse renewal processes have this property, in addition to only having a single state transitioned to on a 1 / 0. Thus, renewal and reverse renewal processes are a subset of alternating renewal processes. Using these rules, we can identify which users' models correspond to renewal, reverse renewal, or alternating renewal processes. We find that 1881 (13.1\%) of the $\epsilon$-machines correspond to (homogeneous) Bernoulli processes, 5408 (37.7\%) correspond to two-state renewal / reverse renewal processes, 2713 (18.9\%) correspond to pure renewal processes with three or more states, 85 (0.59\%) correspond to pure reverse renewal processes with three or more states, and 1250 (8.7\%) correspond to alternating renewal processes with four or more states.


%
%
%




As we have seen, by definition the non-alternating renewal users must be such that their $\epsilon$-machine has one or more states transitioned to on an $x$ from a state transitioned to on an $x' \neq x$. In practical terms, this means that for these users, knowledge of the time since a user switched from a period of activity / quiescence to a period of quiescence / activity is not sufficient to resolve a causal state. However, in many cases it is sufficient to know the behavior of the user immediately prior to a switch from quiescence to activity or vice versa. For example, a user may behave differently when they have switched from active to passive after just being active compared to after just being passive. These cases correspond to generalizations of the alternating renewal process to higher orders. Table~\ref{tab:alternating-renewal-counts} summarizes the number of models that correspond to an alternating renewal model of a certain order. For example, a zeroth order alternating renewal model corresponds to a Bernoulli process, a first order alternating renewal process corresponds to the model architecture in Figure~\ref{Fig-eM-mixed-renewal-architecture}, etc. We see that many of the models resolve to alternating renewal models of higher orders. In total, 95\% of the users have an $\epsilon$-machine in agreement with an alternating renewal model of order 6 or smaller.

\begin{table}
	\centering
	\caption{The number (proportion) of users with $\epsilon$-machines and $\epsilon$-transducers of a given alternating renewal order. A 0th order alternating renewal process corresponds to a Bernoulli process, a 1st order alternating renewal process corresponds to a standard renewal process, etc.}
	\begin{tabular}{c | r | c}
		Alternating Renewal Order & \#  of $\epsilon$Ms & \# of $\epsilon$Ts \\ \hline
		0 & 1881 (13.1\%) & 648 (5.1\%)\\
		1 & 9546 (66.7\%) & 8249 (65.3\%)\\
		2 & 611 (4.3\%) & 400 (3.2\%)\\
		3 & 493 (3.4\%) & 309 (2.4\%)\\
		4 & 530 (3.7\%) & 243 (1.9\%)\\
		5 & 518 (3.6\%) & 220 (0.02\%)\\
		6 & 134 (1.0\%) & 147 (0.01\%)\\
	\end{tabular}
	\label{tab:alternating-renewal-counts}
\end{table}

\subsection{$\epsilon$-transducer Causal Architectures}

Thus far, we have considered the models associated with user behavior when we ignore their inputs. Next we turn to the models that incorporate those inputs, namely the $\epsilon$-transducer. Recall that the input $\{ Y_{t} \}$ we consider is whether a user was mentioned during time interval $t$. As with the $\epsilon$-machines for the self-driven model, we find that most users are well-described by $\epsilon$-transducers with a small number of states, with 95\% having 6 states or fewer and 90\% having 25 or fewer states in the self-memoryless and self-memoryful cases, respectively.

\textbf{Renewal-like Self-Memoryless Transducer.} For the self-memoryless case, 12059 of the 12641 mentioned users (95\%) have an $\epsilon$-transducer with an architecture analogous to Figure~\ref{Fig-eM-renewal-architecture}a. That is, the user has a `just-mentioned' state, and subsequent time steps without the user receiving a mention lead to transitions away from this state, until a terminal state $\tilde{n}$ is reached. The causal states therefore map to the time since the user was mentioned, with all times of length $\tilde{n}$ or longer mapped to the same state. Thus, when viewed as purely socially-driven, the relevant quantity to track for almost all of the users is the time since they were last mentioned.

\begin{figure}
        \centering
                \includegraphics[width=0.5\textwidth]{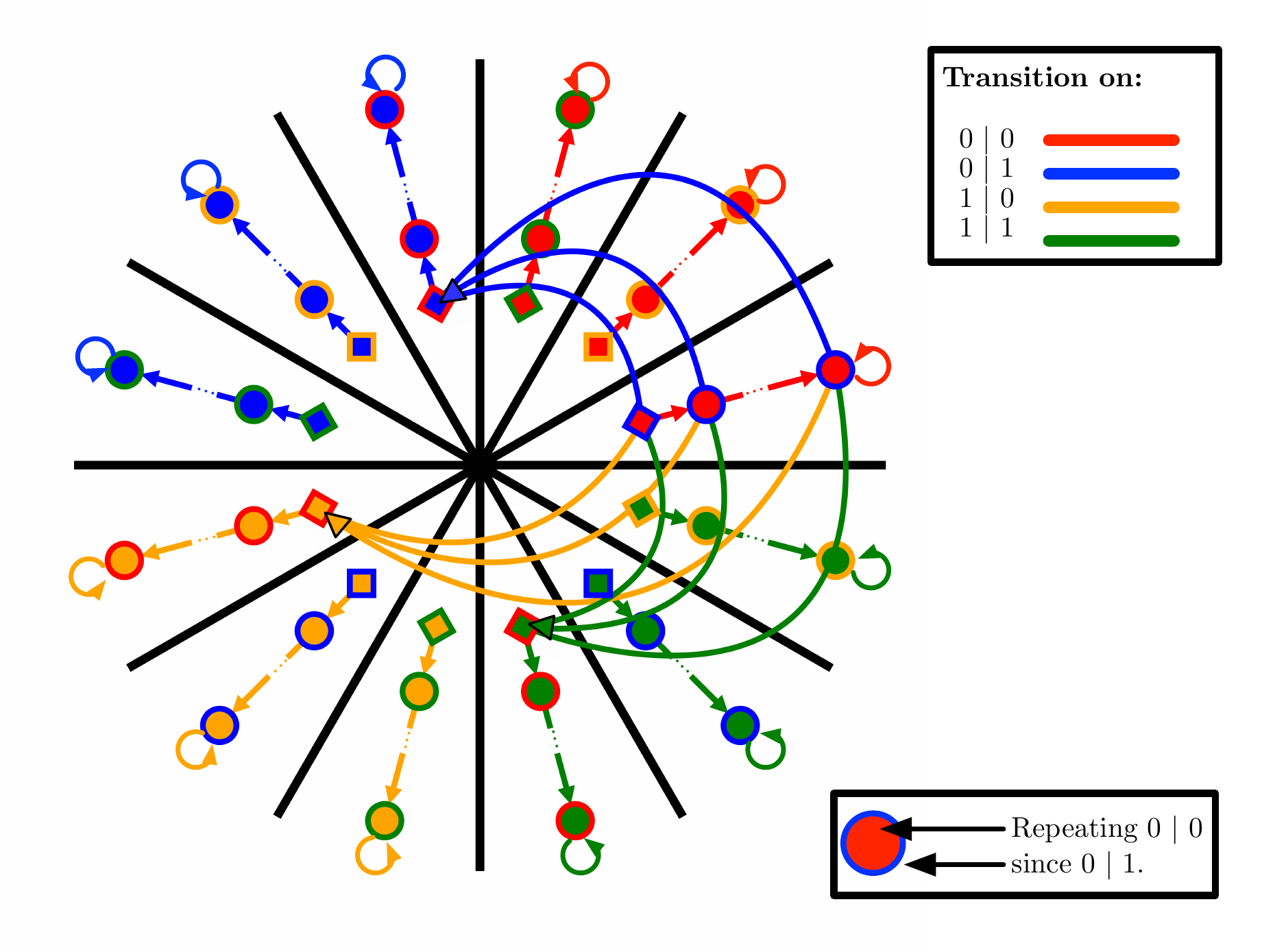}
        \caption{A schematic demonstrating the partitioning of the transducer state space associated with a renewal-like $\epsilon$-transducer. Each quadrant is determined by the input-output symbol pair being `counted,' and each third within a quadrant is determined by the input-output pair the count begins from. We only show outgoing transitions for the first third of the first quadrant, which correspond to transitions of $0 \mid 1$, $1 \mid 0$ or $1 \mid 1$ after observing $0 \mid 0$. 70.4\% of users have $\epsilon$-transducers of this type.}
        \label{Fig-eT-memoryful-architecture}
\end{figure}

\begin{figure}[!h]
	\centering
	\includegraphics[width=0.5\textwidth]{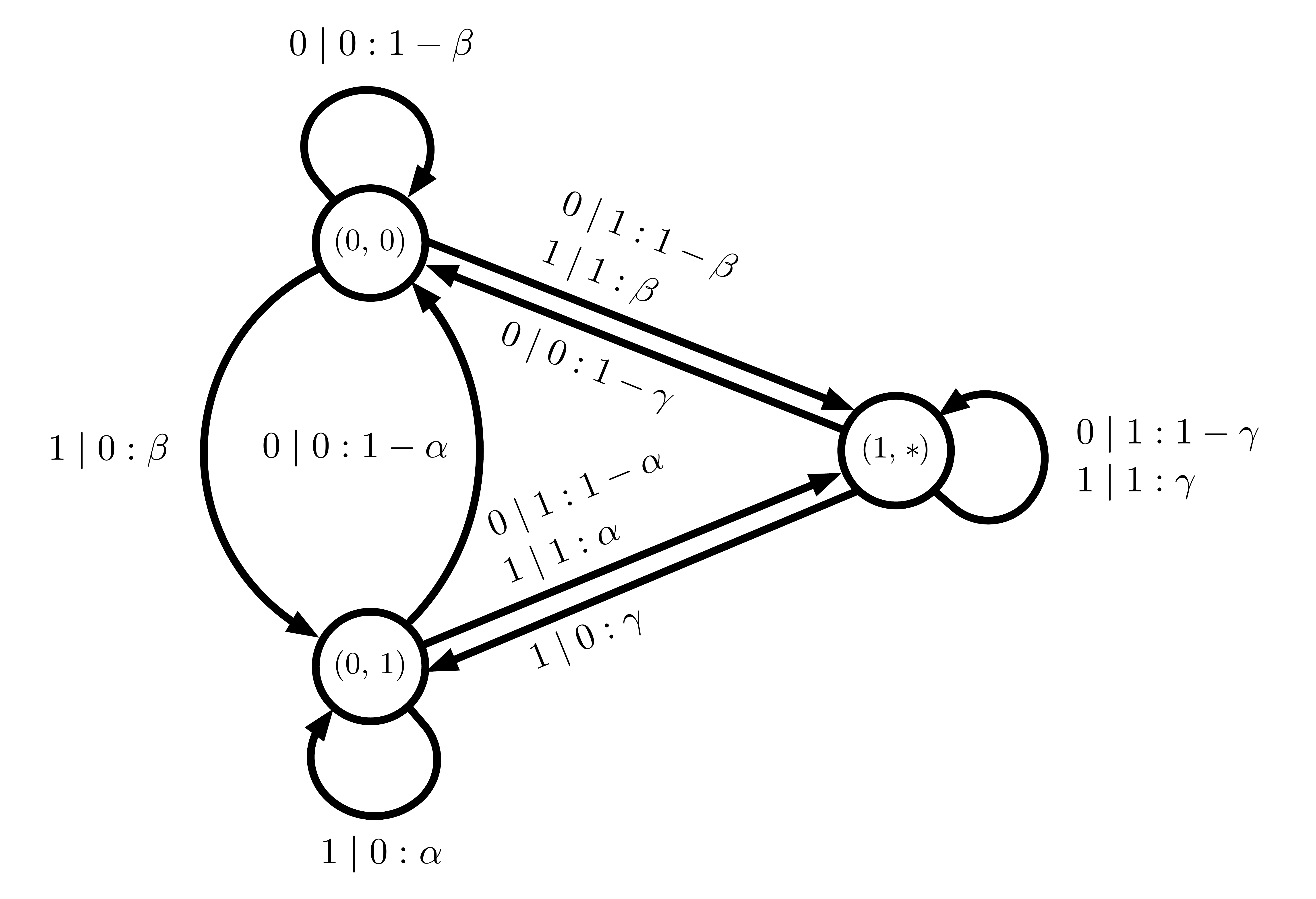}
	\caption{The most common self-memoryful $\epsilon$-transducer architecture associated with 3376 of the 12641 mentioned users (27\%).}
	\label{Fig-social-eT}
\end{figure}

\textbf{Renewal-like Self-Memoryful Transducer.} A similar overarching `counting' model architecture is also present amongst the memoryful \hbox{$\epsilon$-transducers}. Recalling that another way to view the states of an alternating renewal process is as counting the length of runs of $x$ since the last $x' \neq x$, we can generalize this to the \hbox{$\epsilon$-transducer} by considering states that count the lengths of runs of input-output symbols $(y, x)$ since the last input-output symbol $(y', x') \neq (y, x)$. As in the memoryless $\epsilon$-transducer case, we call this an alternating renewal-like process, since the causal states act in a similar fashion. We present a schematic representation of the partitioning of the channel causal state space in Figure~\ref{Fig-eT-memoryful-architecture}. For these $\epsilon$-transducers, the causal states can be partitioned based on the runs of $(y, x)$ they count since the last $(y', x') \neq (y, x)$. Thus, we begin by dividing the set of causal states into four quadrants, based on the runs of $(y, x)$ which they count. All states in a quadrant labeled by $(y, x)$ are transitioned to on $(y, x)$. Then, the causal states within a quadrant are further partitioned into thirds, where each third corresponds to the symbol $(y', x')$ seen before the current run of $(y, x)$. Thus, each third has a unique start state that is transitioned to on a $(y, x)$ from a state transitioned to on a $(y', x')$. 10216 of 12641 (81\%) of the mentioned users have an $\epsilon$-transducer in this alternating renewal-like class. Note that the partitioning given in Figure~\ref{Fig-eT-memoryful-architecture} is the most general possible for this type of $\epsilon$-transducer. The quadrants / thirds within a quadrant may further collapse, as dictated by the structure of the $\epsilon$-transducer. For example, Figure~\ref{Fig-social-eT} is an alternating renewal-like transducer inferred for 27\% of the mentioned users. This $\epsilon$-transducer has three states, which correspond to runs of $(0, 0)$, $(0, 1)$, and $(1, *)$ and are labeled as such. In this case, the quadrants corresponding to $(1, 0)$ and $(1, 1)$ collapse, since the corresponding state counts runs of $y = 1$ regardless of the user behavior $x$. Moreover, all thirds within a given quadrant also collapse, since the states treat runs of $(y, x)$ as the same from any $(y', x') \neq (y, x)$. In terms of the actual behavior of the user, we see that the state labeled $(0, 0)$ corresponds to when the user has been both quiescent and unmentioned in the recent past. In this case, the user has probability $\beta$ of being active given this state. The state labeled $(0, 1)$ corresponds to when the user has been active, but not mentioned, in the recent past. In this case, the user has probability $\alpha > \beta$ of being active given this state. Finally, the state labeled $(1, *)$ corresponds to the case where the user has been mentioned in the recent past, regardless of whether or not the user has been active. The user has probability $\gamma > \beta$ of being active given this state. Thus, for over a quarter of the users, we see that knowledge of the recent past of both their own and their inputs behaviors provides sufficient information for predicting their future behavior. In particular, each of the quadrants requires only a single state, whereas in the most general model of this type with $L_{\max} = 6$ allows for $6\times3 = 18$ states per quadrant.

As with the renewal, reverse renewal, and alternating renewal processes inferred from the $\epsilon$-machines, this alternating renewal-like $\epsilon$-transducer motivates a particular parametric model, albeit a much more complicated one. In this case, we need to specify the chain lengths $\tilde{n}_{(y', x'), (y, x)}$  within each third $(y', x')$ of a quadrant $(y, x)$. However, this results in at most $L \cdot |\mathcal{X}| \cdot |\mathcal{Y}|\cdot (|\mathcal{X}| \cdot |\mathcal{Y}| - 1)$ states overall, compared to $(|\mathcal{X}| \cdot |\mathcal{Y}|)^{L}$ states in the most general model, and therefore a linear growth in model complexity as a function of history length $L$ compared to a geometric growth.

Again, as with the alternating renewal process, the alternating renewal-like $\epsilon$-transducer generalizes to higher orders by considering the input-output behavior immediately prior to a switch from $(y', x')$ to $(y, x) \neq (y', x')$. For example, a second order alternating renewal-like $\epsilon$-transducer would distinguish between a user becoming quiescent and unmentioned after being mentioned twice in the past compared to going unmentioned before the previous mention. Many of the users exhibit $\epsilon$-transducers of higher order as shown in Table~\ref{tab:alternating-renewal-counts}. Of the 12641 mentioned users, 78\% are alternating renewal-like of order 6 or smaller.


\section{Conclusions}

In this paper, we have developed and applied a modeling framework for human behavior in digital environments. The approach begins by viewing a user's behavior as a discrete-time point process at a prespecified temporal resolution, and then considers four possible stochastic models that might give rise to the user's behavior, namely the seasonal, self-driven, socially-driven, and self- and socially-driven processes which we estimate using an inhomogeneous Bernoulli process, an $\epsilon$-machine, or self-memoryless/memoryful $\epsilon$-transducers.

We have found that simple computational architectures, as specified by their $\epsilon$-machines and $\epsilon$-transducers, describe much of the observed behavior of the users in our data set. A renewal process model, or its generalizations to reverse renewal and alternating renewal processes, was found to be appropriate for approximately 80\% of the users in our study. This is in agreement with much of the literature on human communication patterns. However, we emphasize that we did not \emph{assume} such models \emph{a priori}, but rather \emph{discovered} their prevalence by using non-parametric modeling in an exploratory fashion. In fact, the appearance of reverse renewal and alternating renewal processes demonstrate that renewal process models alone are not sufficient to describe, for example, the burstiness observed in human communication patterns.  Moreover, we discovered a new class of renewal-like models that generalize renewal processes to input-output systems. We found that this class of models describes over 70\% of the users in terms of the interaction between their activity and their social inputs. The prevalence of these stereotyped $\epsilon$-machines/transducers motivates the use of either frequentist (such as the cross-validation approach used in this paper) or Bayesian (as recently developed in~\cite{strelioff2014bayesian}) approaches that take advantage of these structures \emph{a priori} during the estimation process. In addition to the generalized alternating renewal models, more general models were necessary for over 20\% of the users in the self-driven case and nearly 30\% of the users in the self- and socially-driven case.

One short-coming of our work is the discretization of the behavior of all users at the time resolution of $\delta = $ 10 minutes in order to analyze their behavior through the lens of discrete-time computational mechanics. The computational mechanics of continuous-time, discrete event processes was recently developed in~\cite{marzen2017informational,marzen2017structure}. While at present machine reconstruction algorithms for inferring $\epsilon$-machines from such processes do not exist, this work and others motivate their development. A reanalysis of this data set from this viewpoint may reveal additional details hidden by the discretization.

The apparent complexity of observed user behavior on Twitter seems to arise from a simple computational landscape. Our present work lays out an initial sketch of its features. We hope this work motivates further exploration of the computational landscape of user behavior on Twitter and other communication platforms, and refinement of their maps.

\appendix

\section{The \texttt{transCSSR} Algorithm for $\epsilon$-transducer Reconstruction}

\label{transCSSR-algorithm}

A diverse collection of algorithms have been developed to infer $\epsilon$-machines from data, from the topological methods first presented in~\cite{crutchfield1989inferring} to more recent methods based on Bayesian methods~\cite{strelioff2014bayesian}. Additional algorithms have been developed based on spectral methods~\cite{varn2013machine} and integer programming~\cite{paulson2014computational}. We focus on the Causal State Splitting Reconstruction (\texttt{CSSR}) algorithm~\cite{CSSR-UAI-2004}.

The theory formalizing $\epsilon$-transducers has only recently been developed, and the literature on $\epsilon$-transducer reconstruction from finite data is sparse. Sketches of \texttt{CSSR}-like algorithms for $\epsilon$-transducer reconstruction are provided in~\cite{shalizi2001computational,haslinger2010computational}. In this appendix, we develop the ideas originally suggested in these prior works, and present a generalization of \texttt{CSSR} for $\epsilon$-transducer reconstruction from data resulting from input-output systems. In homage to \texttt{CSSR}, we call our algorithm \texttt{transCSSR}, a portmanteau of transducer and \texttt{CSSR}. The \texttt{transCSSR} algorithm has been implemented in Python, and is available on GitHub~\cite{transcssr2015}.

We sketch the \texttt{transCSSR} algorithm here, and give the pseudocode in Figure~\ref{fig:transCSSR-pseudocode}. The first phase of the algorithm groups input-output histories into a set of weakly prescient states. It begins by assuming that all input-output histories induce the same one-step-ahead predictive distribution. This is equivalent to assuming the transducer output future is independent of the input-output past, or to grouping together all joint histories into a candidate causal state represented by the joint input-output suffix $(*\lambda, *\lambda)$, where $\lambda$ is the null symbol. At each successive step, each candidate causal state $s$ is tested for weak prescience by growing its histories into the past by one input-output pair. If the state is weakly prescient, then the predictive distribution for the new history should be equivalent to that of its parent causal state. This condition is tested using the null hypothesis
\begin{align}
\begin{split}
&P\left(X_{0} \mid (Y_{-(L+1)}^{-1}, X_{-(L+1)}^{-1}) = (a y_{-L}^{-1}, b  x_{- L}^{-1})\right) \\&= P\left(X_{0} \mid \hat{S} = s\right) \label{null-hyp}
\end{split}
\end{align}
with a significance test of size $\alpha$. If the null hypothesis is rejected, the predictive distribution of the history is compared against all of the remaining candidate causal states $s^{*} \neq s$ using the restricted alternative hypothesis
\begin{align}
\begin{split}
&P\left(X_{0} \mid (Y_{-(L+1)}^{-1}, X_{-(L+1)}^{-1}) = (a y_{-L}^{-1}, b  x_{- L}^{-1})\right) \\ &=P\left(X_{0} \mid  \hat{S} = s^{*}\right). \label{restricted-alt-hyp}
\end{split}
\end{align}
Finally, if the history's predictive distribution does not agree with any of the candidate causal states, it is split into a new candidate casual state. Such potential splitting is performed until the input-output symbols under consideration are each of length $L_{\max}$. Any hypothesis test for comparing discrete distributions may be used for (\ref{null-hyp}) and (\ref{restricted-alt-hyp}). We use the test based on the $G$-statistic~\cite{harremoes2012information}.

At the end of the this stage, each candidate causal state consists of histories that are within-state equivalent and between-state distinct in terms of their predictive distributions, and each candidate causal state is weakly prescient. The causal states have these properties, in addition to being deterministic / unifilar on transitions between states on input-output pairs. To ensure determinism / unfilarity, the successor state for each history $(\mathbf{y}, \mathbf{x})$ of length $L_{\max} - 1$ on an input-output pair $(a, b) \in \mathcal{Y} \times \mathcal{X}$ is determined. If two or more histories in the same state transition to different states on the input-output pair $(a, b)$, that state is split, and the determinization step is repeated. This procedure is repeats until all transitions are deterministic. Since there are finitely many histories, this procedure always terminates, in the extreme case with each history having its own causal state. Because this stage only ever splits histories from states, and the states before this stage were weakly prescient and within-state equivalent and between-state distinct, the resulting states are as well. Therefore, the procedure results in a set of states that are weakly prescient and deterministic, and thus causal.

\makeatletter
\newcommand*{\rom}[1]{\expandafter\@slowromancap\romannumeral #1@}
\makeatother

\newlength\myindent
\setlength\myindent{2em}
\newcommand\bindent{%
  \begingroup
  \setlength{\itemindent}{\myindent}
  \addtolength{\algorithmicindent}{\myindent}
}
\newcommand\eindent{\endgroup}

\begin{figure}[!h]
	\begin{spacing}{0.8}
\begin{algorithmic}
\STATE{\rom{1}. Initialization: $L \leftarrow 0$, $\Sigma \leftarrow \{ \{ \emptyset\} \}$}
\STATE{\rom{2}. Homogenization:}
\bindent
\WHILE{$L < L_{\text{max}}$}
\FOR{each $s \in \Sigma$}
\STATE{estimate $\hat{P}(X_{0} | \hat{S} = s)$}
\FOR{each $(\mathbf{y}, \mathbf{x}) \in s$ }
\FOR{each $(a, b) \in \mathcal{Y} \times \mathcal{X}$}
\STATE{estimate \\$p \leftarrow$ \\ $\hat{P}(X_{0} | (Y_{-L-1}^{-1}, X_{-L-1}^{-1}) = (a \mathbf{y}, b \mathbf{x}))$}
\STATE{\textsc{Test}$(\Sigma, p, (a \mathbf{y}, b \mathbf{x}), s, \alpha)$}
\ENDFOR
\ENDFOR
\ENDFOR
\STATE{$L \leftarrow L + 1$}
\ENDWHILE
\eindent
\STATE{\rom{3}. Determinization:}
\bindent
\STATE{Remove transient states from $\Sigma$}
\STATE{\texttt{recursive} $\leftarrow$ \textsc{False}}
\WHILE{\textsc{Not} \texttt{recursive}}
\STATE{\texttt{recursive} $\leftarrow$ \textsc{True}}
\FOR{each $s \in \Sigma$}
\FOR{each $(a, b) \in \mathcal{Y} \times \mathcal{X}$}
\STATE{$(\mathbf{y}_{0},\mathbf{x}_{0}) \leftarrow$ first $(y_{- (L_{\max} - 1)}^{-1}, x_{- (L _{\max} - 1)}^{-1}) \in s$}
\STATE{$T(s, (a, b)) \leftarrow \hat{\epsilon}((\mathbf{y}_{0}a, \mathbf{x}_{0}b))$}
\FOR{each $(\mathbf{y}, \mathbf{x}) \in s, (\mathbf{y}, \mathbf{x}) \neq (\mathbf{y}_{0}, \mathbf{x}_{0})$}
\IF{ $\hat{\epsilon}((\mathbf{y}a, \mathbf{x}b)) \neq T(s, (a, b))$}
\STATE{create new state $s' \in \Sigma$}
\STATE{$T(s', (a, b)) \leftarrow \hat{\epsilon}((\mathbf{y}a, \mathbf{x}b))$}
\FOR{each $(\mathbf{y}', \mathbf{x}') \in s$ such that \\ $\hat{\epsilon}((\mathbf{y}'a, \mathbf{x}'b)) = \hat{\epsilon}((\mathbf{y}a, \mathbf{x}b))$}
\STATE{\textsc{Move}$((\mathbf{y}', \mathbf{x}'), s, s')$}
\ENDFOR
\STATE{\texttt{recursive} $\leftarrow$ \textsc{False}}
\ENDIF
\ENDFOR
\ENDFOR
\ENDFOR
\ENDWHILE
\eindent
\STATE
\STATE{\textsc{Test}$(\Sigma, p, (a \mathbf{y}, b\mathbf{x}), s, \alpha)$}
\bindent
\IF{null hypothesis (\ref{null-hyp}) passes a test of size $\alpha$}
\STATE{$s \leftarrow \{(a \mathbf{y}, b\mathbf{x})\} \cup s $}
\ELSIF{restricted alternative hypothesis (\ref{restricted-alt-hyp}) passes a test of size $\alpha$ for $s^{*} \in \Sigma, s^{*} \neq s$}
\STATE{\textsc{Move}$((a \mathbf{y}, b\mathbf{x}), s, s^{*})$}
\ELSE
\STATE{create new state $s' \in \Sigma$}
\STATE{\textsc{Move}$((a \mathbf{y}, b\mathbf{x}), s, s')$}
\ENDIF
\eindent
\STATE
\STATE{\textsc{Move}($(\mathbf{y}, \mathbf{x}), s_{1}, s_{2})$}
\bindent
\STATE{$s_{1} \leftarrow s_{1} \setminus \{ (\mathbf{y}, \mathbf{x})\}$}
\STATE{re-estimate $\hat{P}(X_{t} | \hat{S} = s_{1})$}
\STATE{$s_{2} \leftarrow s_{2} \cup \{ (\mathbf{y}, \mathbf{x})\}$}
\STATE{re-estimate $\hat{P}(X_{t} | \hat{S} = s_{2})$}
\eindent
\end{algorithmic}
\end{spacing}
\caption{Pseudo-code for the \texttt{transCSSR} algorithm. Arguments: $\mathcal{Y}, \mathcal{X}$: the discrete alphabets for the input and output processes; $(y_{1}^{N}, x_{1}^{N})$: the joint input-output sequence of length $N$ drawn from $\mathcal{Y}\times\mathcal{X}$; $L_{\max}$: the maximum history length used when estimating candidate causal states; $\alpha$: the probability of falsely rejecting the null hypotheses (\ref{null-hyp}) and (\ref{restricted-alt-hyp}).}
\label{fig:transCSSR-pseudocode}
\end{figure}


%

\end{document}